\documentclass[twocolumn,amsmath,amssymb,superscriptaddress,aps,prl]{revtex4-1}

\usepackage[dvips]{epsfig}
\usepackage{graphicx}% Include figure files
\usepackage{amsmath}
\usepackage{amssymb}
\usepackage{commath} %http://ctan.mackichan.com/macros/latex/contrib/commath/commath.pdf
\usepackage{xspace}
\usepackage{fancyhdr}
\usepackage{afterpage}
\usepackage{xcolor}
\usepackage{dcolumn}
\usepackage{comment}
\usepackage{listings}
\usepackage{makeidx}
\usepackage{longtable}
\usepackage{bm}

\newcommand{\rw}{\ensuremath{R_w}\xspace}

\newcommand{\sjba}[1]{}

\newcommand{\mta}[1]{}

\definecolor{dgreen}{HTML}{008000}

\definecolor{codegreen}{rgb}{0,0.6,0}
\definecolor{codegray}{rgb}{0.5,0.5,0.5}
\definecolor{codepurple}{rgb}{0.58,0,0.82}
\definecolor{backcolour}{rgb}{0.95,0.95,0.92}
\lstdefinestyle{mystyle}{
    backgroundcolor=\color{backcolour},
    commentstyle=\color{codegreen},
    keywordstyle=\color{magenta},
    numberstyle=\tiny\color{codegray},
    stringstyle=\color{codepurple},
    basicstyle=\ttfamily\footnotesize,
    breakatwhitespace=false,
    breaklines=true,
    captionpos=b,
    keepspaces=true,
    numbers=left,
    numbersep=5pt,
    showspaces=false,
    showstringspaces=false,
    showtabs=false,
    tabsize=2
}
% ...
% ...
% ...
\newcounter{saveenumi}

\newcommand{\be}{\begin{enumerate}}
\newcommand{\ee}{\end{enumerate}}
\newcommand{\bes}{\begin{enumerate}[wide, labelwidth=!, labelindent=0pt, label=\textbf{\textcolor{blue}{\arabic*}.}]}
\newcommand{\ees}{\end{enumerate}}

\def\snse{SnSe\xspace}
\def\rrangee{$10 \le r \le 60$~\AA\xspace}
\def\trangex{$300 \le  T \le 1070$~K\xspace}
\def\trangen{$300 \le  T \le 975$~K\xspace}
\def\rrange{$1.5 \le r \le 10$~\AA\xspace}

\graphicspath{{./}{Figures//}}

\def\cmcm{Cmcm\xspace}
\def\pnma{Pnma\xspace}
\def\t2g{$t_{2g}$\xspace}
\def\ts{T$_{s}$\xspace}
\def\rw{r$_{w}$\xspace}

\begin{document}
%
%%%%%%%%%%%%%%%%%%%%%%%%%%%%%%%%%%%%%%%%%%%%%%%%%%%%%%%%%%%%%%%%
% Abstract
%%%%%%%%%%%%%%%%%%%%%%%%%%%%%%%%%%%%%%%%%%%%%%%%%%%%%%%%%%%%%%%%
%
\begin{abstract}
The local atomic structure of \snse\ was characterized across its orthorhmbic-to-orthorhombic structural phase transition using x-ray pair distribution function analysis.
Substantial Sn off-centering distortions persist in the high symmetry high temperature phase, with symmetry different from that of ordered distortions below the transition.
The analysis implies that the transition is neither order-disorder nor displacive, but rather a complex crossover where the character of coupling changes from 3D-like at low temperature to 2D-like at high temperature.
Robust ferro-coupled SnSe intra-layer distortions suggest a ferroelectric-like instability as the driving force.
Complex local Sn off-centering is integral to the ultra-low lattice thermal conductivity mechanism in \snse.
\end{abstract}
%%%%%%%%%%%%%%%%%%%%%%%%%%%%%%%%%%%%%%%%%%%%%%%%%%%%%%%%%%%%%%%

\title{Nanoscale Sn off-centering behind low thermal conductivity in {SnSe} thermoelectric}

%%%%%%%%%%%%%%%%%%%%%%%%%%%%%%%%%%%%%%%%%%%%%%%%%%%%%%%%%%%%%%%
\author{E.~S.~Bozin}\email[]{bozin@bnl.gov}
\affiliation{Condensed Matter Physics and Materials Science Division, Brookhaven National Laboratory, Upton, NY 11973, USA}
\author{H.~Xie}
\affiliation{Department of Chemistry, Northwestern University, Evanston, Illinois 60208, USA}
\author{A.~M.~M.~Abeykoon}
\affiliation{Photon Sciences Division, Brookhaven National Laboratory, Upton, New York 11973, USA}
\author{S.~M.~Everett}
\affiliation{Neutron Scattering Division, Oak Ridge National Laboratory, Oak Ridge, Tennessee 37831, USA}
\author{M.~G.~Tucker}
\affiliation{Neutron Scattering Division, Oak Ridge National Laboratory, Oak Ridge, Tennessee 37831, USA}
\author{M.~G.~Kanatzidis}
\affiliation{Department of Chemistry, Northwestern University, Evanston, Illinois 60208, USA}
\author{S.~J.~L.~Billinge}
\affiliation{Condensed Matter Physics and Materials Science Division, Brookhaven National Laboratory, Upton, NY 11973, USA}
\affiliation{Department of Applied Physics and Applied Mathematics, Columbia University, New York, NY~10027, USA}
%%%%%%%%%%%%%%%%%%%%%%%%%%%%%%%%%%%%%%%%%%%%%%%%%%%%%%%%%%%%%%%

\date{\today}
\maketitle

Suppression of heat transport is critical~\cite{nolas;mrsb06} to high performance thermoelectricity~\cite{mahan;pnas96}, requiring the ability to effectively scatter  heat-bearing phonons~\cite{sales;mrsb98,nolas;arms99,snyde;nm04,morel;prl08,miyat;sa17,vones;prl17,zhao;afm20}.
This can be accomplished either via crystal engineering~\cite{hsu;s04,liu;am17,vones;nm13,zheng;aem15,chen;nc17}, or by utilization of materials with intrinsically low lattice thermal conductivity~\cite{liu;nm12,takab;rmp14}.
In the latter case, particularly intriguing are systems exhibiting nanoscale symmetry-broken
%emergent on heating from undistorted high-symmetry ground
states that are intrinsic and driven by electronic instabilities.
For example, in thermoelectric PbTe such a state emerges on warming  and involves Pb off-centering displacements~\cite{bozin;s10} and associated giant anharmonic phonon scattering~\cite{delai;nm11}.
The stereochemical activity of 6s$^{2}$ electron lone pairs results in the formation of local dynamic correlated dipoles~\cite{jense;prb12,niel;ees13,sangi;prm18}.
A similar emergent state (emphanisis) was found in {AgGaTe${_2}$}, where local off-centering of Ag is driven by a weak sd$^{3}$ orbital hybridization, resulting in strong {acoustic-optical} phonon scattering and an ultralow lattice thermal conductivity~\cite{xie;am22}.
These examples show the diversity of electronic instabilities that can lead to hidden nanostructural responses important for applications.

\begin{figure}[tb]
\includegraphics[width=0.475\textwidth]{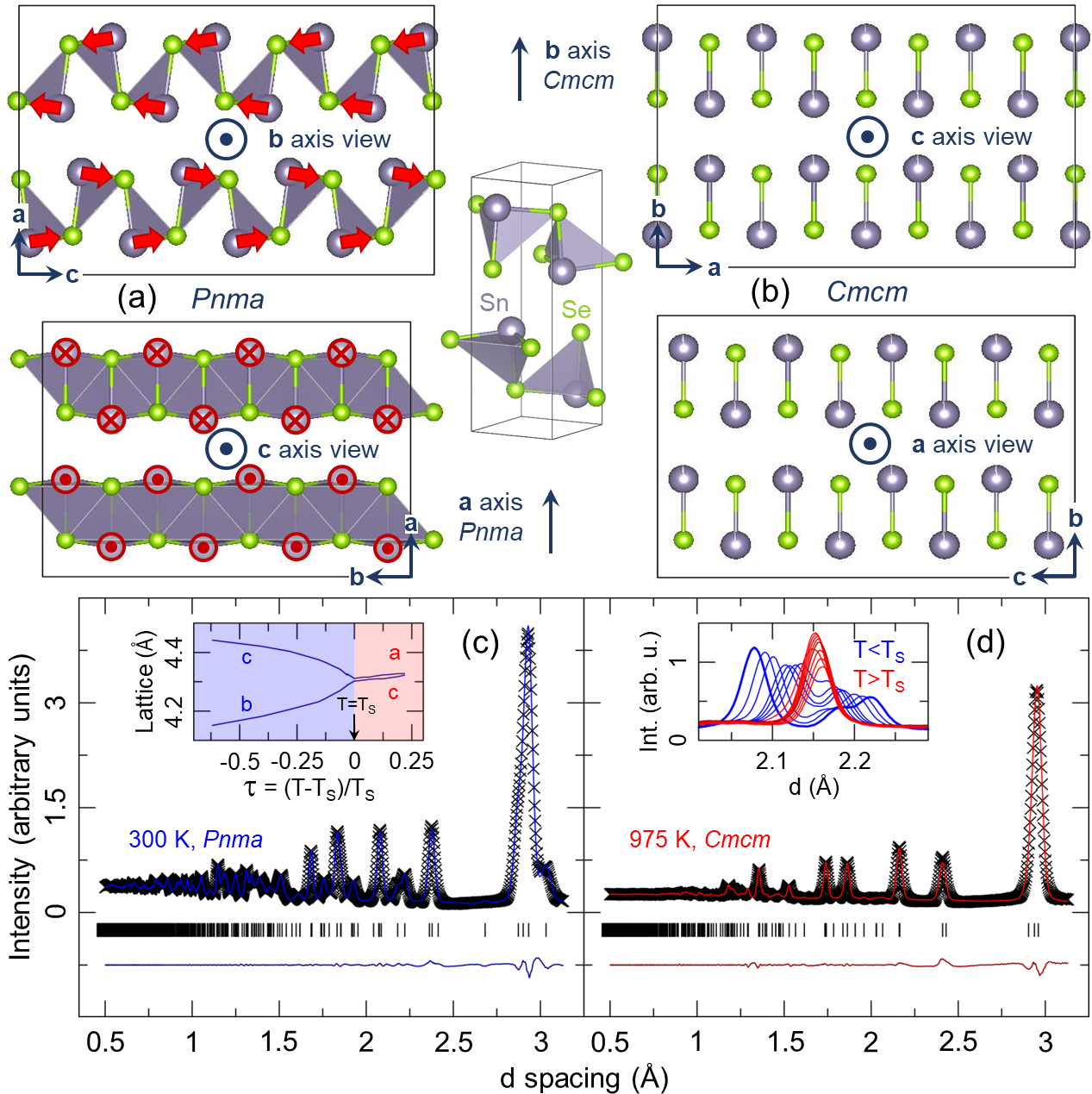}
\caption{\label{fig:structure_properties}
Crystal structure of \snse\ and temperature induced phase transition.
 The structure features two SnSe bilayers which are related by inversion symmetry.
(a) Low temperature \pnma\ phase. Sn is in off-centered, forming strongly bonded SnSe$_{3}$ pyramids.
The associated atomic displacements exhibit intra-bilayer "ferro" and inter-bilayer "antiferro" ordering, as indicated by red arrows.
(b) High temperature \cmcm\ phase. Sn is centered, resulting in an Sn-Se dumbbell motif when viewed down the $c$-axis, being the shortest bond in the structure.
To enable comparison in (a) and (b), an arbitrary stack of unit cells, depicted in the middle for \pnma, along the bilayer directions is shown.
Rietveld fits of neutron powder diffraction data confirm the average sample structure across the transition: \pnma\ at 300~K (c) and \cmcm\ at 975~K (d).
Evolution of the in-plane lattice parameters and the \cmcm\ (200)/(002) reflection across the transition ($T_{s}\approx 840$~K) are shown in the insets to (c) and (d).
}
\end{figure}

Here we address the binary semiconductor \snse, Fig.~\ref{fig:structure_properties}(a)~\cite{okaza;jpsj56}, a novel high efficiency high temperature thermoelectric exhibiting ultralow thermal conductivity and high thermoelectric figure of merit ZT~\cite{zhao;n14,zhao;s16}.
Due to its remarkable optoelectronic properties, \snse\ has also been extensively explored for photovoltaic applications~\cite{butt;cec14,zhao;nr15,liu;cc15,reddy;jmsme16,shi;ads18}, whereas its monolayer variants feature robust and tunable ferroelectricity~\cite{chang;nl20,higas;nc20,orlov;epl21,zhu;jap21}.

The material has a rather simple structure featuring two-dimensional (2D) bilayers.
At low temperature it forms an orthorhombic  structure in the \pnma\ space group.
It has a structural phase transition to a higher symmetry but also orthorhombic (\cmcm) structure,  Fig.~\ref{fig:structure_properties}(b), at $T_{s}\approx 800$~K~\cite{wiede;zk79,chatt;jpcs86,sist;acb16}.
As with the above examples, a complex interplay of lattice and electronic degrees of freedom are at play, in this case involving 5s$^{2}$ Sn electron lone pairs~\cite{zhao;ees16}, multiple band edges~\cite{bansal;prb16,siddi;acsami22}, and highly anisotropic bonding~\cite{zhou;nm21}.
A large lattice anharmonicity exhibited by the high-energy optical phonons and driven by a ferroelectric-like instability~\cite{li;np15,sanch;apl15} has been observed.
Recent first-principles calculations~\cite{hong;mtp19} suggested that a Jahn-Teller-like instability of the band structure causes a lattice distortion in the vicinity of the structural transition, whose coupling to lattice dynamics results in the observed strong anharmonicity and the ultralow thermal conductivity.  We investigate this further using the atomic pair distribution function (PDF) technique~\cite{egami;b;utbp12} a probe of local structure.

%%%%%%%%%%%%%%%%%
% Begin Figure 01
%%%%%%%%%%%%%%%%%
%%%%%%%%%%%%
% End Figure
%%%%%%%%%%%%

We find that, although the average crystallographic structure has Sn ions centered in a distorted square based pyramid of Se ions in the high symmetry \cmcm\ phase, the local structure displays a complex evolution of off-centering distortions that persist on a nanometer length scale up to 1070~K, the highest measured temperature.
The phase transition cannot be regarded as order-disorder, since the local distortions at high $T$ are of different symmetry and magnitude than those observed below \ts.
The analysis suggests a change in local interactions and distortion coupling from 2D-like at $T > T_{s}$ to 3D-like at $T < T_{s}$.

%%%%%%%%%%%%%%%%%
% Begin Figure 02
%%%%%%%%%%%%%%%%%
\begin{figure*}
\includegraphics[width=0.9\textwidth]{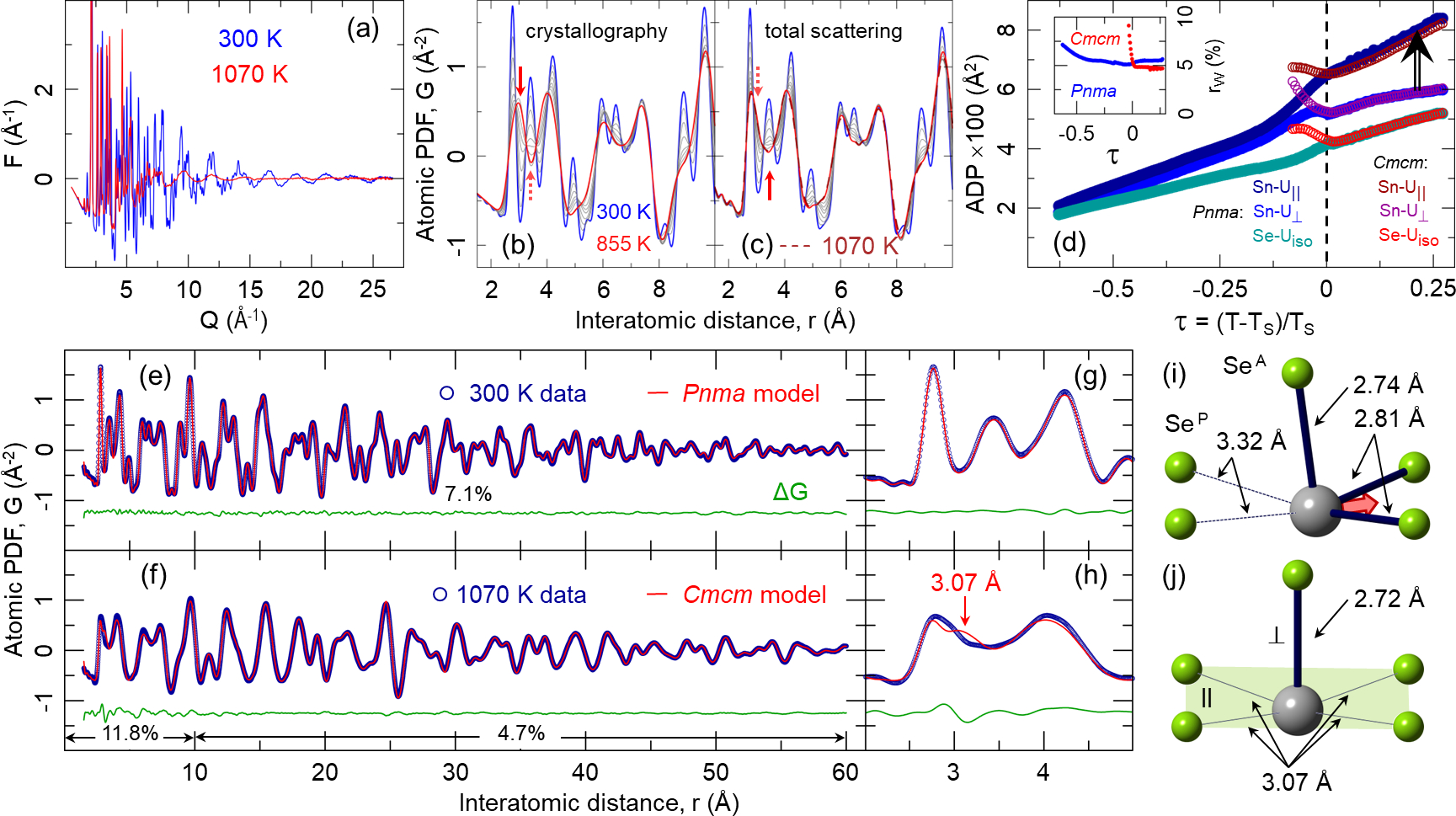}
\caption{\label{fig:structure_lengthscales}
Assessment of diverse structural length scales in \snse.
(a) Measured reduced x-ray total scattering function, F(Q), at 300~K (blue) and 1070~K (red).
Representative PDFs from different temperatures are shown in (b), simulated using structure reported from high resolution x-ray powder diffraction~\cite{sist;acb16}, and (c) measured PDFs from this work.
The PDFs from the extremal temperatures are plotted in red (hot) and blue (cool) colors.
Intermediate temperatures are shown in gray.
Solid/dotted red arrows indicate presence/absence of features discussed in text.
Measured local structure does not change observably in the \cmcm\ phase, between 855~K and 1070~K.
(d) Temperature evolution of atomic displacement parameters (ADPs) obtained from \pnma\ and \cmcm\ PDF models over the temperature range \rrangee .
Abscissa is the reduced temperature, $\tau$.
Models utilized anisotropic Sn ADPs with components parallel and perpendicular to the layers, and isotropic Se ADP, as indicated.
A sizeable enhancement of the in-plane ADP component, marked by a black arrow, is observed at $T > T_{s}$, indicative of planar disordering of the Sn.
The inset shows the fit residual, r$_{w}$, for the two models.
Fits of crystallographic models extended over the entire data range for 300~K and 1070~K are shown in (e) in (f), respectively.
Difference traces, $\Delta$G, in green are offset for clarity.
In panels (g) and (h) the same fits are shown with a focus on the local $r$-range.
An observable discrepancy is seen between the \cmcm\ model and 1070~K data on a sub-nanometer length scale, resulting in an increase of r$_{w}$.
The Sn environments expected from crystallography are depicted in (i) and (j).
In the \pnma\ structure the Sn is off-center, whereas in the \cmcm\ average structure Sn is in the centre of the Se$_{4}$ plaquette.
Labels A and P refer to apical and in-plane Se, respectively.
The arrow in (h) marks a feature expected in the \cmcm\ symmetry, associated with the Sn centering, which is clearly absent in the data.
}
\end{figure*}
%%%%%%%%%%%%
% End Figure
%%%%%%%%%%%%

High quality polycrystalline \snse\ sample was synthesized using the approach reported in Ref.~\onlinecite{zhou;nm21}. It behaves canonically as determined by an average structure analysis of the sample through Rietveld refinements~\cite{toby;jac13} of neutron powder diffraction data collected at beamline BL-1B at the Spallation Neutron Source over \trangen\ range in steps of $\Delta T = 25$~K.
The Bragg analysis confirmed \pnma\ and \cmcm\ structures below and above \ts, respectively,  Fig.~\ref{fig:structure_properties}(c), (d).

The \pnma\ crystal structure contains two \snse\ bilayers in the unit cell, featuring intra-bilayer ferro-ordered Sn displacements and exhibiting inter-bilayer antiferro-coupling, as shown in Fig.~\ref{fig:structure_properties}(a).
The primary component of the distortion is a Sn off-centering from the center of the square formed by its almost coplanar Se ions (Fig.~\ref{fig:structure_lengthscales}(i)).
In the \cmcm\ average structure, the distortions disappear, Fig.~\ref{fig:structure_properties}(b), and Sn is centered in the square with four equal bond-lengths to its coplanar Se ions (Fig.~\ref{fig:structure_lengthscales}(j)).

In the Rietveld refinements of the \cmcm\ phase, we found substantially enlarged and nearly isotropic planar components of Sn's atomic displacement parameters (ADPs).
At 975~K, $U_{11}= 0.075(2)$~\AA$^{2}$ and $U_{33}= 0.078(3)$~\AA$^{2}$, whereas $U_{22} =  0.040(3)$~\AA$^{2}$ in the direction perpendicular to the bilayers.
This is consistent with a high resolution powder diffraction report~\cite{sist;acb16} and indicative of intra-bilayer structural disorder at high T, suggestive of an order-disorder character of the structural transition that should be evident in the local structure.

For the local structural studies we carried out x-ray total scattering experiments in the range \trangex\  in steps of $\Delta T = 5$~K at the 28-ID-1 (PDF) beamline at the National Synchrotron Light Source~II using a 74.5~keV energy beam ($\lambda = 0.1665$~\AA, $Q_{max} = 25$~\AA$^{-1}$) in rapid acquisition mode~\cite{chupa;jac03} using a 2D PerkinElmer area detector.
Finely pulverized \snse\ powder was loaded in an evacuated and flame-sealed quartz capillary with inner diameter of 1~mm.
Sample temperature control on warming was achieved using a FMB Oxford Hot Air Blower model GSB1300.
The PDFs, $G(r)$, were obtained using standard protocols~\cite{juhas;jac13} with the help of the xPDFsuite program~\cite{yang;arxiv14} and modelled over different $r$-ranges using the PDFgui program~\cite{farro;jpcm07}.

To see how the PDFs would behave based on the average crystal structure through the phase transition, in Fig.~\ref{fig:structure_lengthscales}(b) we show PDFs computed from the Rietveld models at selected temperatures below and above \ts~\cite{sist;acb16}.
The actual measured PDFs are shown in Fig.~\ref{fig:structure_lengthscales}(c).
The observed changes in the PDFs with temperature are similar in the two cases.
However, the changes are different in detail in the region of the PDF, $2.8 \le r \le 3.3$~\AA, that describes the Sn-Se planar bonds.
In the average structure, this peak shifts in position to higher-$r$ as the sample goes through the phase transition.
This is because the Sn off-centering reduces in amplitude with increasing temperature, disappearing completely for $T > T_{s}$.
The solid red arrow at $\sim 3.1$~\AA\ in Fig.~\ref{fig:structure_lengthscales}(b) indicates this.
Simultaneously, the peaks at $r \sim 2.8$~\AA\ and at $\sim 3.4$~\AA\ reduce strongly in intensity.
This response in the PDF computed from the average structure comes from Sn-Se planar bonds going from 2-short, 2-long in the \pnma\ structure to 4-medium ($\sim 3.1$~\AA) bonds in \cmcm.
This is shown schematically in Figs.~\ref{fig:structure_lengthscales}(i) and ~\ref{fig:structure_lengthscales}(j).

This is not seen in the measured total scattering PDFs.
The peaks at $r \sim 2.8$ and 3.3~\AA\ broaden with increasing temperature, but they do not shift and no additional intensity grows up at the average $r\sim 3.1$~\AA\  position. The local environment of the Sn does not change significantly with temperature from the \pnma\ phase all the way up to 1070~K (dashed dark red trace in Fig.~\ref{fig:structure_lengthscales}(c)).
We next turn to modeling to quantify and characterize this behavior.

%%%%%%%%%%%%%%%%%
% Begin Figure 03
%%%%%%%%%%%%%%%%%
\begin{figure}[tb]
\includegraphics[width=0.475\textwidth]{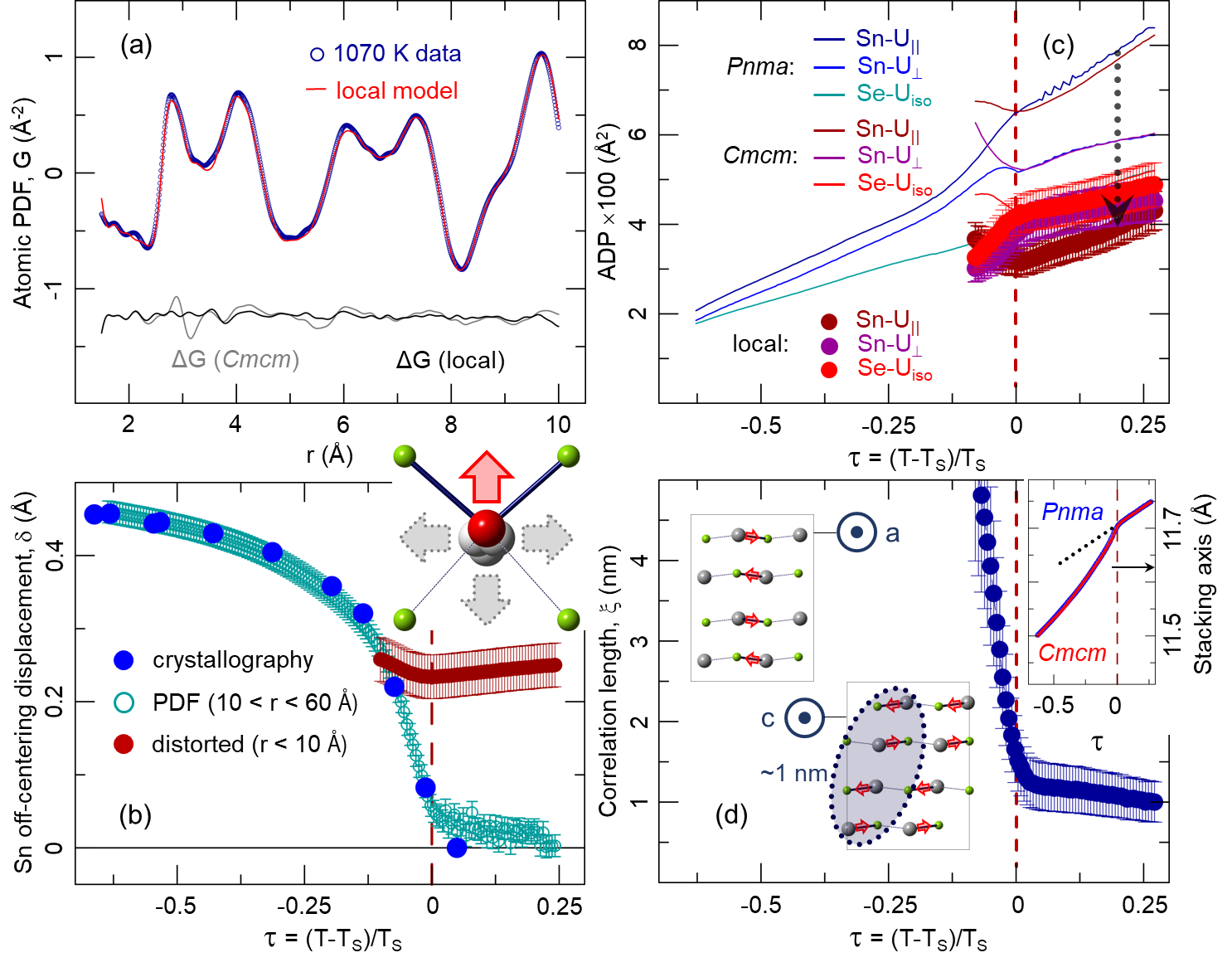}
\caption{\label{fig:structure_offcentering}
Local symmetry breaking Sn off-centering in \snse\ at $T > T_{s}$.
(a) The fit to 1070~K data of the local distorted model featuring split Sn sites, as shown in the inset to (b).
Corresponding difference trace, $\Delta$G (black), is offset for clarity.
The gray difference trace is that of the \cmcm\ model, Fig.~\ref{fig:structure_lengthscales}, is shown as a reference.
(b) The magnitude the Sn off-centering displacements vs. reduced temperature, as seen by high resolution powder diffraction~\cite{sist;acb16}, \rrangee\ range PDF \pnma\ model fits, and the local split site model shown in the inset.
The Se$_{4}$ plaquette is nearly a square in \cmcm\, and the local model allows for displacements in 4~directions indicated by block-arrows.
(c) The distorted model results in ADP magnitudes substantially reduced as compared to these observed by undistorted models.
The ADPs obtained from \pnma\ and \cmcm\, shown in Fig.~\ref{fig:structure_lengthscales}(d), are reproduced as solid lines.
(d) The estimated correlation length of local distortions, as described in the text.
Insets to the left: the displacement coupling, shown in a 2 $\times$\ 1 $\times$\ 2 \cmcm\ supercell, of the best model where the couplings are explicitly tested.
Inset to the right: temperature dependent lattice parameter along the bilayer stacking direction (b axis of \cmcm\ model in red, a axis of \pnma\ model in blue).
Note the change in slope at temperature \ts.
}
\end{figure}
%%%%%%%%%%%%
% End Figure
%%%%%%%%%%%%

The effect observed in the qualitative analysis is clear also in the modeling.
At low temperature, the \pnma\ model fits well over the entire range of the PDF, as evidenced by the small agreement factor (7.1~\%) and the lack of structure in the residual plotted in green and labeled $\Delta G$ in Fig.~\ref{fig:structure_lengthscales}(e).
On the other hand, at high temperature in the \cmcm phse, fitting the \cmcm\ model to the measured PDF results in a very good fit in the high-$r$ region ($> 10$~\AA), but the fit becomes poor in the region below 10~\AA.
The average \cmcm\ structure is not describing the local structure well in the \cmcm\ phase.
The nature of the failure is highlighted in Fig.~\ref{fig:structure_lengthscales}(h) and
indeed, there is a lack of intensity at the average Sn-Se square-planar distance of 3.07~\AA.
In the local structure the Sn is still off-center in its square of Se ions even in the \cmcm\ phase.

The temperature dependence of the offsets can be seen in fits of Sn ADPs of \pnma\ and \cmcm\ models over the \rrangee\ range, shown in Fig.~\ref{fig:structure_lengthscales}(d).
The ADP of Se does not behave anomalously and is almost linear all the way into the \cmcm\ phase.
The component of the Sn ADP perpendicular to the layers is also well-behaved.  However, the in-plane component of the Sn ADP is higher than the other two and increases as the sample passes into the \cmcm\ phase, again supporting the idea that the Sn is off-center in its square-plane of Se ions.
This is consistent with a previously reported gradual deviation of ADPs from these predicted by the Debye model as the transition is approached on warming~\cite{sist;acb16}.

The simplest explanation for the observed behavior is that the phase transition has an order-disorder character and the \pnma\ distortions survive to high-$T$ but become disordered among the different variants.
More detailed modeling can establish if this is indeed the case.
Different distorted local models were initially fit to the 1070~K data over the narrow $r$-range (\rrange) where the misfits in the \cmcm\ model are seen.

The model featuring \pnma\ symmetry with ferro-coupled displacements along the $b$-axis and within the bilayer, Fig.~\ref{fig:structure_properties}(a), resulted in an improved fit residual compared to \cmcm\ (7.5~\% vs. 11.8\%).
However, this model did not fully account for the observed misfit seen in Figs.~\ref{fig:structure_lengthscales}(f,h) and returned a dramatically enlarged intra-layer ADP in the direction perpendicular to the Sn displacement (U$_{33}$ = 0.16~\AA).
This is inconsistent with isotropic intra-layer ADPs observed in broad range fits described above, implying that there is a distortion component in the \cmcm\ phase also along the $c$-axis, which is not accounted for by the \pnma\ symmetry.
Constraining the intra-layer ADPs to be isotropic resulted in similar issues.
This is not a purely order-disorder transition.

A split site model depicted in the inset to Fig.~\ref{fig:structure_offcentering}(b) was introduced next, in which isotropy of intra-layer ADPs was maintained.
In this, each of the crystallographic Sn and Se sites within the \cmcm\ symmetry was split into four equally populated sites (25~\% occupancy each).
Each site of the same type (Sn or Se) in the model was allocated the same displacement parameter away from the centered \cmcm\ configuration and along both $a$ and $c$ crystallographic directions.
This model resulted in a fit with appreciably improved residual (5.8~\%), Fig.~\ref{fig:structure_offcentering}(a), and significantly reduced ADPs, Fig.~\ref{fig:structure_offcentering}(c).
Notably, the observed values of ADPs at high temperature agree well with the Debye model prediction~\cite{sist;acb16}.
The model yielded Sn displacements, $\delta$, of $\sim 0.25$~\AA, and an order of magnitude smaller Se displacements ($\sim 0.02$~\AA).
This suggests dynamic motions of displaced Sn ions in something reminiscent of a Mexican hat potential occurring in the \cmcm phase, similar to the high temperature phase of hexagonal manganites~\cite{skjae;prx19}.
This model was then applied to all high temperature data.
Nearly temperature-independent Sn displacements ensued, as shown in Fig.~\ref{fig:structure_offcentering}(b).
On the other hand, the fit of the split site model dramatically deteriorated below \ts implying that the local distortion symmetry starts changing at the transition.
This is also revealed by the distortion magnitude displaying an apparent increase at $T < T_{s}$.

The temperature dependence of the spatial correlations of the local distortions, $\xi$, was estimated from the $r$-dependent (cumulative) integrals of the absolute value of the difference between the data and the \cmcm\ model, $\lvert \Delta G(r)\rvert$, over the full range of data (60~\AA).
The characteristic length scale of short-range correlations is defined as the value of $r$ at which the integral departs from a linear trend associated with statistical noise in the PDF data~\cite{bozin;sr14}.
In Fig.~\ref{fig:structure_offcentering}(d), $\xi(T)$ increases slowly on decreasing temperature for $T > T_{s}$, but exhibits a rapid increase at $T < T_{s}$.
This behavior is highly unusual.
If the local distortions were related to critical fluctuations of the \pnma\ phase, it would be expected that their correlation length diverges at \ts as this is approached from above.
This is clearly not the case, as $\xi(T)$ exhibits a rapid upturn at and \emph{below} \ts.
The distortions associated with disorder at $T > T_{s}$ are therefore likely not directly related to the instability driving the \pnma\ ordering, further corroborating the conclusion that the transition is not of conventional order-disorder type.

Models incorporating different combinations of displacement patterns involving two bilayers have also been tested against the high temperature data.
In these models, displacements were allowed to be ferro- or antiferro-coupled within individual layers of the bilayers, with inter-bilayer ferro- or antiferro-coupling.
These short range models resulted in fits of similar quality (5.5~\%$ < r_{w}  < 6.8$~\%).
The model involving a pattern with ferro-coupled distortions within individual sub-layers within the bilayer, and antiferro-coupling between the sub-layers in the bilayer, sketched in the inset to Fig.~\ref{fig:structure_offcentering}(d), resulted in the lowest \rw , with fit residual and ADPs comparable to those observed by the split site model (5.5~\%).
This may suggest that, on a nanometer length scale at high temperature, the ferro-coupling is 2D in nature (with antiferro, or possibly disordered, intra-bilayer correlations).  This is in contrast to the robust 3D-like ferro intra-bilayer coupling observed in the \pnma\ phase.  It is possible that the origin of the phase transition is a 2D-3D crossover, where somewhat robust fluctuating 2D locally-ferro-coupled dipoles can only order into a static structure when a much weaker perpendicular coupling is able to drive the ordering.
This dimensionality crossover might be further supported by the observation of a dramatic change in the temperature dependence of the lattice parameter in the stacking direction at the transition (upper right inset to Fig.~\ref{fig:structure_offcentering}(d)).
The behavior is seen whether the data are modeled with the \pnma\ or the \cmcm\ models and it shows a rapid expansion along the stacking direction in the ordered \pnma\ phase but a much slower expansion above the transition.

Another notable observation is that the anomalous upturn in ADPs on warming starts at $\sim 600$~K~\cite{sist;acb16}, well below the phase transition (-0.25 in reduced temperature units), suggesting that the dynamic disorder is setting in even while there remains a net long-range ordered off centered distortion with the \pnma\ character.
The average \pnma\ off-centering distortion amplitude is falling off rapidly in this region and there may be a competition between 2D fluctuating dipoles and the full 3D order and the presence or absence of an average long-range ordering depends on whether the ferro order can percolate.
It is in this region close to and above the structural transition where the thermal conductivity of the material becomes very low~\cite{zhou;nm21}.

In summary, we have shown that, in the vicinity of and above the phase transition where the thermoelectricity is very good, \snse\ is nanostructured with electronically driven intrinsic fluctuating Sn off-centering distortions.  These are short-range correlated with ferro-ordering preferred within a single layer within the bilayers but with an apparent loss of coherence of the dipole ordering along the stacking direction even within the bilayer.

%
%%%%%%%%%%%%%%%%%%%%%%%%%%%%%%%%%%%%%%%%%%%%%%%%%%%%%%%%%%%%%%%%
% Acknowledgments
%%%%%%%%%%%%%%%%%%%%%%%%%%%%%%%%%%%%%%%%%%%%%%%%%%%%%%%%%%%%%%%%
%
\begin{acknowledgments}
Work at Brookhaven National Laboratory was supported by U.S. Department of Energy, Office of Science, Office of Basic Energy Sciences (DOE-BES) under contract No. DE-SC0012704.
MGK acknowledges partial support from the U.S. Department of Energy, Office of Science Basic Energy Sciences under grant DE-SC0014520, DOE Office of Science (thermoelectric investigations).
X-ray PDF measurements were conducted on beamline 28-ID-1 of the National Synchrotron Light Source II, a U.S. Department of Energy (DOE) Office of Science User Facility operated for the DOE Office of Science by Brookhaven National Laboratory under Contract No. DE-SC0012704.
Neutron powder diffraction data were collected at BL-1B beamline of the Spallation Neutron Source, a U.S. Department of Energy Office of Science User Facility operated by the Oak Ridge National Laboratory.
\end{acknowledgments}

%%%%%%%%%%%%%%%%%%%%%%%%%%%%%%%%%%%%%%%%%%%%%%%%%%%%%%%%%%%%%%%%%%
%\bibliography{snse-emil}
%\bibliographystyle{apsrev}
%\vfill\newpage
%\renewcommand\thefigure{S\arabic{figure}}
%\setcounter{figure}{0}
%\section{Supplementary Information}

%%%%%%%%%%%%%%%%%%
%% Begin Figure S01
%%%%%%%%%%%%%%%%%%
%\begin{figure}[tpb]
%\includegraphics[width=3.5in]{resist.pdf}
%\caption{\label{fig:transport} Temperature dependence of resistivity, $\rho$, of \fese\ sample.
%Relevant temperatures $T_s$ and $T_c$ are labeled and marked with arrows.}
%\end{figure}
%%%%%%%%%%%%%
%% End Figure
%%%%%%%%%%%%%

\end{document}